\documentclass[
superscriptaddress,
final,
reprint,
showkeys,
pre,
aps,
floatfix,
]{revtex4-2}

\usepackage[dvips]{graphicx}
\usepackage[utf8]{inputenc}
\usepackage{amsmath,amssymb,booktabs,latexsym,MnSymbol,bm}
\usepackage[T1]{fontenc}

\usepackage[%
  colorlinks=true,
  urlcolor=blue,
  linkcolor=blue,
  citecolor=blue
]{hyperref}
\usepackage[final]{pdfpages}
\makeatletter
\AtBeginDocument{\let\LS@rot\@undefined}
\makeatother
\begin{document}

\title{Interpretable liquid crystal phase classification via two-by-two ordinal patterns}

\author{Leonardo G. J. M. Voltarelli} 
\affiliation{Departamento de F\'isica, Universidade Estadual de Maring\'a, Maring\'a, PR 87020-900, Brazil}

\author{Natalia Osiecka-Drewniak} 
\author{Marcin Piwowarczyk} 
\author{Ewa~Juszyńska-Gałązka} 
\affiliation{The Henryk Niewodniczański Institute of Nuclear Physics, Polish Academy of Sciences, E. Radzikowskiego 152, 31-342 Kraków, Poland}

\author{Rafael~S.~Zola} 
\affiliation{Departamento de F\'isica, Universidade Estadual de Maring\'a, Maring\'a, PR 87020-900, Brazil}
\affiliation{Departamento de F\'isica, Universidade Tecnol\'ogica Federal do Paran\'a, Apucarana, PR 86812-460, Brazil}

\author{Matja{\v z} Perc} 
\affiliation{Faculty of Natural Sciences and Mathematics, University of Maribor, Koro{\v s}ka cesta 160, 2000 Maribor, Slovenia}
\affiliation{Community Healthcare Center Dr. Adolf Drolc Maribor, Ulica talcev 9, 2000 Maribor, Slovenia}
\affiliation{Department of Physics, Kyung Hee University, Dongdaemun-gu, Seoul 02447, Republic of Korea}
\affiliation{University College, Korea University, Seongbuk-gu, Seoul 02841, Republic of Korea}

\author{Haroldo V. Ribeiro} 
\email{hvribeiro@uem.br}
\affiliation{Departamento de F\'isica, Universidade Estadual de Maring\'a, Maring\'a, PR 87020-900, Brazil}

\date{\today}

\begin{abstract}
Liquid crystal textures encode rich structural information, yet mapping these images to mesophase identity remains challenging because visually similar patterns can arise from distinct structures. Here we present a simple, interpretable representation that maps textures to a 75-dimensional frequency vector of two-by-two ordinal patterns, grouped into eleven symmetry-based types to characterize a large-scale dataset spanning seven mesophases. Combined with a simple machine learning classifier, this lightweight representation yields near-perfect phase recognition, including the difficult distinction between smectic A and smectic B mesophases. Our approach generalizes to unseen compounds and accurately distinguishes between phase identity and material origin. Unlike deep learning methods, each ordinal pattern is readily interpretable, and model explanations augmented with network visualizations of pattern interactions reveal the specific types and pairwise dependencies that drive each mesophase decision, providing compact, physically meaningful summaries of texture determinants. These results establish two-by-two ordinal patterns as an interpretable and scalable tool for liquid crystal image analysis, with potential applications to other complex patterned systems in materials science.
\end{abstract}

\maketitle

\section*{Introduction}

Liquid crystals, intermediate states of matter between crystalline solids and isotropic liquids, exhibit characteristic anisotropic and viscoelastic properties~\cite{collings2019introduction} that underpin their widespread use in display technologies~\cite{goodby2014handbook} and other advanced applications~\cite{kato2018functional, bisoyi2021liquid, yin2022advanced}. Under polarized light microscopy, liquid crystals display a wide range of optical textures~\cite{dierking2003textures, dierking2025liquid}: colorful interference patterns arising from birefringence and spatial variations in the average molecular alignment. These textures encode rich structural information, and their appearance depends on the liquid crystal phase, molecular orientation, sample preparation, and the presence of topological defects; yet distinct structural arrangements can yield visually similar patterns, making interpretation non-trivial~\cite{dierking2003textures}. The information-rich and visually complex nature of liquid crystal textures, together with the rapid adoption of machine learning in the physical sciences~\cite{baldi2014searching, mukund2017transient, carleo2019machine, dreissigacker2019deep}, has motivated growing efforts to apply data-driven approaches to analyze liquid crystal images and their associated physical properties~\cite{piven2024machine}. Much of this work has focused on classifying mesophases directly from optical textures~\cite{sigaki2020learning, dierking2023classification, dierking2023testing, dierking2023deep, betts2023machine, osiecka2023distinguishing, osiecka2024liquid, osiecka2024machine, betts2024possibilities, terroa2025convolutional}, predominantly using deep learning methods, while other studies have explored applications ranging from predicting nematic elastic constants~\cite{zaplotnik2023neural} and dielectric behavior~\cite{taser2022comparison} to detecting structural features in smectic films~\cite{hedlund2022detection} and identifying defects~\cite{kang2009automatic, minor2020end, liu2011automatic}.

In parallel to these developments, but in contrast to the often black-box nature of deep learning, ordinal methods have emerged as a promising framework for investigating liquid crystal textures. These approaches are based on the seminal work of Bandt and Pompe on permutation entropy~\cite{bandt2002permutation} and its subsequent extensions to two-dimensional data~\cite{ribeiro2012complexity, zunino2016discriminating, pessa2020mapping}. Applied to liquid crystal textures, ordinal analysis has proven effective in extracting relevant physical parameters (\textit{e.g.}, nematic order parameter, sample temperature, cholesteric pitch length)~\cite{sigaki2019estimating}, classifying phase transitions, predicting critical temperatures, and distinguishing among different doping concentrations~\cite{pessa2022determining}. The recently introduced nearest-neighbor permutation entropy~\cite{voltarelli2024characterizing}, which calculates permutation entropy from random walks on a graph of spatially proximate pixels, has achieved an accuracy comparable to that of deep learning approaches in predicting the cholesteric pitch length from liquid crystal textures.

Despite these advances, most studies have considered a small set of mesophases, leaving the optical and structural diversity of liquid crystal textures still underexplored. Prevailing machine learning approaches also often lack interpretability, obscuring how model parameters are learned and what drives predictions. These limitations highlight the need for scientifically grounded methods that use universally applicable, interpretable, and yet sufficiently simple descriptors to enable testable predictions and rigorous validation~\cite{amaral2024artificial}, particularly in liquid crystal image analysis~\cite{piven2024machine}. Furthermore, previous ordinal approaches for analyzing liquid crystal textures do not explicitly handle identical pixel intensities, instead resolving these ties with random or deterministic heuristics -- a practice long known to cause information loss in time-series analysis~\cite{zunino2017permutation, bian2012modified} and only recently examined in the context of images~\cite{tarozo2025two}. Here, we address these issues by investigating a large-scale dataset comprising over $13{,}000$ textures obtained from several compounds exhibiting seven liquid crystal phases. To do so, we analyze the ordering patterns of pixel intensities within two-by-two partitions that traverse the textures one pixel at a time~\cite{bandt2023two}, while explicitly accounting for identical pixel intensities~\cite{tarozo2025two}. 

Our procedure maps every texture to a 75-dimensional space defined by the frequencies of ordinal patterns, which we further group into eleven categories based on continuity and symmetry. The resulting features are readily interpretable -- capturing sharp edges, corner-like motifs, uniform versus textured regions, and vertical or horizontal structures -- and support expert visual exploration, showing that similar textures tend to cluster in this ordinal-pattern space (see an interactive visualization in Ref.~\cite{voltarelli2026umaplc}). Indeed, textures from each phase exhibit distinct distributions of ordinal patterns, enabling near-perfect phase classification using a method considerably simpler and lighter than deep neural networks. Furthermore, we show that our ordinal approach effectively distinguishes between the highly similar textures of the smectic A and smectic B phases~\cite{dierking2003textures, osiecka2023distinguishing}, significantly outperforming baseline classifiers, even when applied to textures from compounds not included in the training data. Beyond accurate predictions, combining our ordinal representation with explainable machine learning analyses reveals the key ordinal patterns -- and their interactions -- that characterize each phase. 

\section*{Materials and Methods}

\subsection*{Experimental setup and data collection}

We begin by describing the dataset of liquid crystal textures used in this work, which is an adaptation of the collection introduced by Osiecka-Drewniak \textit{et al.}~\cite{osiecka2024liquid}. Our study focuses on seven distinct phases exhibited by these compounds: nematic, smectic A (SmA), smectic C (SmC), and crystalline-like smectic phases B (SmB), F (SmF), G (SmG), and I (SmI). These textures were obtained using a polarizing microscope (Leica DM2700P) with samples mounted on a heating stage (Linkam THMS600) and precise temperature regulation via a controller (Linkam T96-S). The samples were placed between glass plates and heated above the isotropization temperature. The images were collected during cooling from 120$^\circ$C to 0$^\circ$C at a rate of 10$^\circ$C/min. To reduce potential biases due to non-uniform illumination, we followed standard microscopy practices aimed at producing a uniform and reproducible illumination field during image acquisition. In each imaging session, samples were carefully aligned with the optical axis of the microscope, illumination and condenser settings were kept fixed, and the acquisition software automatically adjusted the exposure time to maintain consistent brightness and contrast across images. Although sample thickness was not explicitly controlled, textures were acquired at multiple positions and orientations to mitigate potential residual illumination asymmetries. The nematic, SmC, SmF, SmG, and SmI phases were observed in compounds from three homologous series: 7OABOOCx (for x = 1-11; (E)-4-[(4-heptyloxyphenyl)diazenyl]phenyl alkanoates), 8OABOOCy (for y = 2, 8; 12, (E)-4-[(4-octyloxyphenyl)diazenyl]phenyl alkanoates), and 10OABOOCz (for z = 1-11; (E)-4-[(4-decanyloxyphenyl)diazenyl]phenyl alkanoates)~\cite{piwowarczyk2024azobenzene, piwowarczyk2025terminal}. The SmA and SmB phases were observed in  8BBAA (4-bromobenzylidene-4$'$-octyloxyaniline) and 9BBAA (4-bromobenzylidene-4$'$-nonyloxyaniline)~\cite{osiecka2023distinguishing}. In these two compounds, the SmA-SmB transition corresponds specifically to the transition from fluid SmA to soft crystal B previously characterized in Ref.~\cite{osiecka2023distinguishing}. Figure~S1 in the Supplemental Material~\cite{SI} depicts the structural formula of these compounds.

The final dataset comprises $13{,}601$ PNG images at $600\times 600$ pixels and 24-bit color (8 bits per RGB channel). Five phases -- nematic, SmC, SmF, SmG, and SmI -- are each represented by $2{,}000$ images, while SmA and SmB phases contain $1{,}687$ (828 from 8BBAA and 859 from 9BBAA) and $1{,}914$ (951 from 8BBAA and 963 from 9BBAA) images, respectively. For analysis, the three color layers were converted to grayscale using the standard luminance transform~\cite{scikit-image}, yielding a matrix of pixel intensities computed as $0.2125R + 0.7154G + 0.0721B$, where $R$, $G$, and $B$ denote the red, green, and blue channels.

\subsection*{Two-by-two ordinal patterns}

\begin{figure*}[!ht]
  \centering
  \includegraphics[width=.82\linewidth]{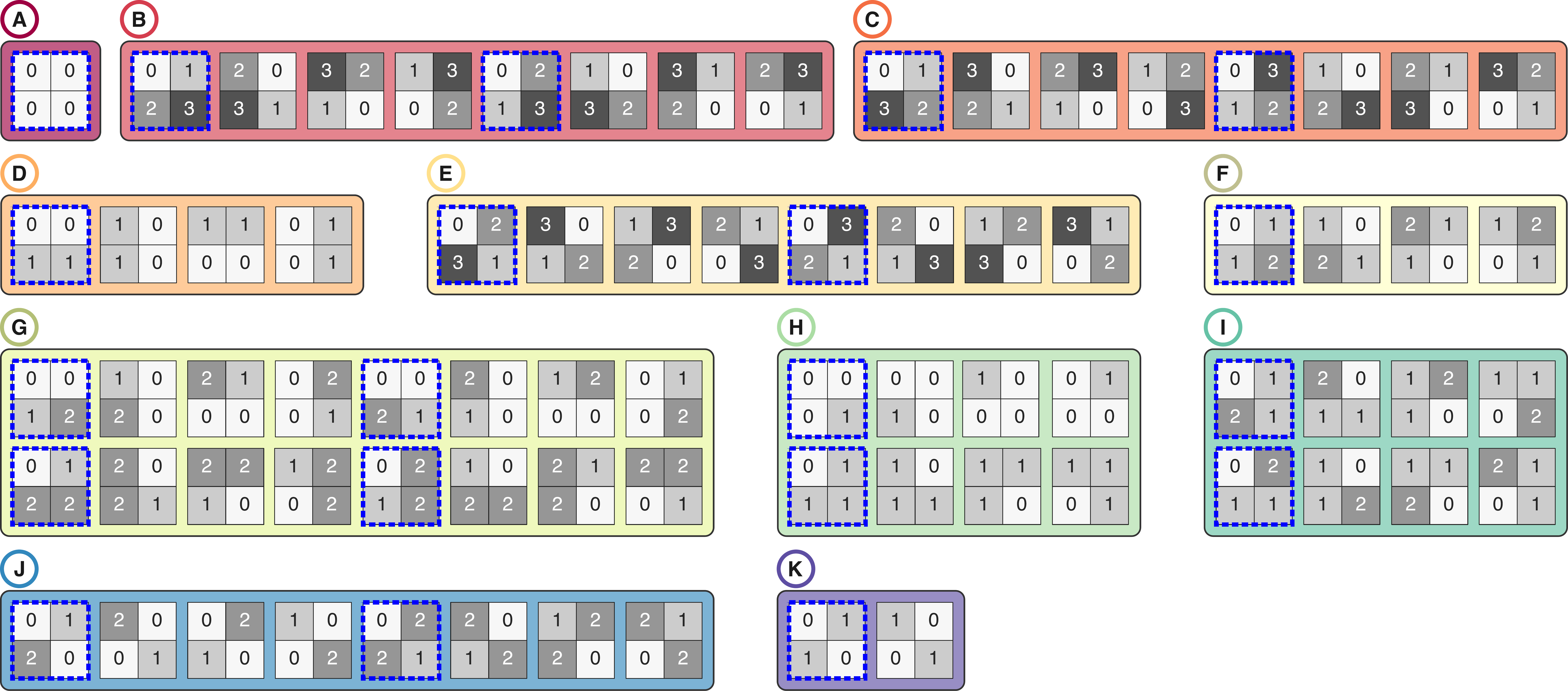}
  \caption{Visualization of the $75$ possible two-by-two ordinal patterns. Patterns are aggregated into 11 groups (labeled with capital letters from A to K) according to their number of unique symbols (distinct intensity levels) and smoothness degree~\cite{tarozo2025two}. Each two-by-two matrix represents one ordinal pattern, where the entries are ranks of pixel intensities within an image partition. For example, the single type A pattern, $[0000]$, corresponds to a uniform partition where all pixels have the same intensity, whereas the first type B pattern, $[0123]$, indicates that pixel intensities increase in rank from top-left to bottom-right within an image partition. Dashed outlines highlight the primary patterns whose rotations by quarter, half, and three-quarter turns yield all remaining patterns of the corresponding group.}
  \label{fig:1}
\end{figure*}

To characterize the liquid crystal textures, we employ the two-by-two ordinal-pattern framework recently introduced for the analysis of paintings~\cite{tarozo2025two}. This approach is based on the Bandt-Pompe formalism~\cite{bandt2002permutation} and its two-dimensional generalization~\cite{ribeiro2012complexity, zunino2016discriminating}, while extending the Bandt–Wittfeld classification of two-by-two ordinal patterns~\cite{bandt2023two} to explicitly represent equal-intensity pixels, an essential feature for robust classification of painting styles~\cite{tarozo2025two}. Assuming that each texture is represented as a matrix $A$ with entries $a_{ij}$ denoting the intensity of the pixels in the row $i$ and the column $j$, we sample $A$ using a sliding $2 \times 2$ window $w$ that traverses the entire image one pixel at a time in both horizontal and vertical directions. At each position, we flatten the submatrix into a vector $w=(a_0,a_1,a_2,a_3)$ in row-major order and replace the pixel values with their ranks. This ranking process explicitly handles ties by assigning the same rank to equal intensities, thereby defining the ordinal pattern of the local pixel arrangement. For instance, $w=(1,2,4,2)$ yields the ordinal pattern $[0121]$, where the first element is assigned rank $0$ for being the smallest, the second and fourth elements share rank $1$ for being the second-largest, and the third element is assigned rank $2$ for being the largest.

A total of $75$ distinct ordinal patterns account for all possible rank combinations, including those with ties. These patterns can be further aggregated into $11$ groups based on the number of unique symbols (distinct intensity levels) and their smoothness degree, as detailed Ref.~\cite{tarozo2025two}. Figure~\ref{fig:1} shows all possible two-by-two ordinal patterns arranged by group. Within each group, labeled with capital letters from A to K, patterns outlined by dashed lines are the primary patterns, whose rotations by quarter, half, and three-quarter turns generate all other patterns within each group. Group A comprises the single pattern $[0000]$, representing a uniform partition of four identical intensity levels. The two-symbol patterns, found in groups D, H, and K, represent partitions with pairs or trios of identical values arranged along rows, corners, or diagonals. Three-symbol patterns are differentiated into four groups (F, G, I, and J) based on the position and relative rank of a single pair of identical values. Finally, the four-symbol patterns, which have no ties and are equivalent to types I, II, and III of Ref.~\cite{bandt2023two}, are classified into groups B, C, and E. These groups describe distinct spatial gradients, ranging from monotonic intensity changes in both directions (B), to patterns with a monotonic trend along one axis but a mixed trend along the other (C), and finally patterns with mixed trends in both directions (E).

For each texture, we therefore evaluate the distribution of ordinal patterns $P=\{p_i;~i=1,\dots,75\}$, where $p_i$ denotes the relative frequency of the $i$-th pattern, ordered as in Figure~\ref{fig:1} (for example, $p_1$ corresponds to $[0000]$, $p_2$ to $[0123]$, and $p_3$ to $[2031]$). This procedure maps all information encoded in the $360{,}000$ pixels of each texture to a $75$-dimensional feature vector that characterizes the local spatial structure of pixel intensities and reflects distinct textural features, such as uniformity, sharp edges, and directional structures. Crucially, because this representation is built from local rank relations among neighboring pixels rather than absolute intensity values, it is invariant under monotonic intensity transformations~\cite{ribeiro2012complexity, zunino2016discriminating} and therefore largely independent of overall brightness and contrast. The local nature of ordinal patterns also makes them robust to common experimental illumination artifacts, such as gradual shading: as long as these effects appear as low-frequency variations across the field of view and do not become strong at the pixel scale, they have little impact on the ordering relations among adjacent pixels. Summing the relative frequencies of patterns belonging to the same group yields a more compact representation of textures in $11$ dimensions that preserves essential ordinal structures, with its components corresponding to the probability of finding ordinal patterns from groups A to K. We rely on the Python module ordpy~\cite{pessa2021ordpy} for all ordinal-pattern computations. In line with observations for art paintings~\cite{tarozo2025two}, we expect textures from distinct mesophases to exhibit characteristic ordinal-pattern distributions and to occupy different regions of this feature space, enabling effective characterization and classification.

\begin{figure*}[!ht]
    \centering
    \includegraphics[width=.9\linewidth]{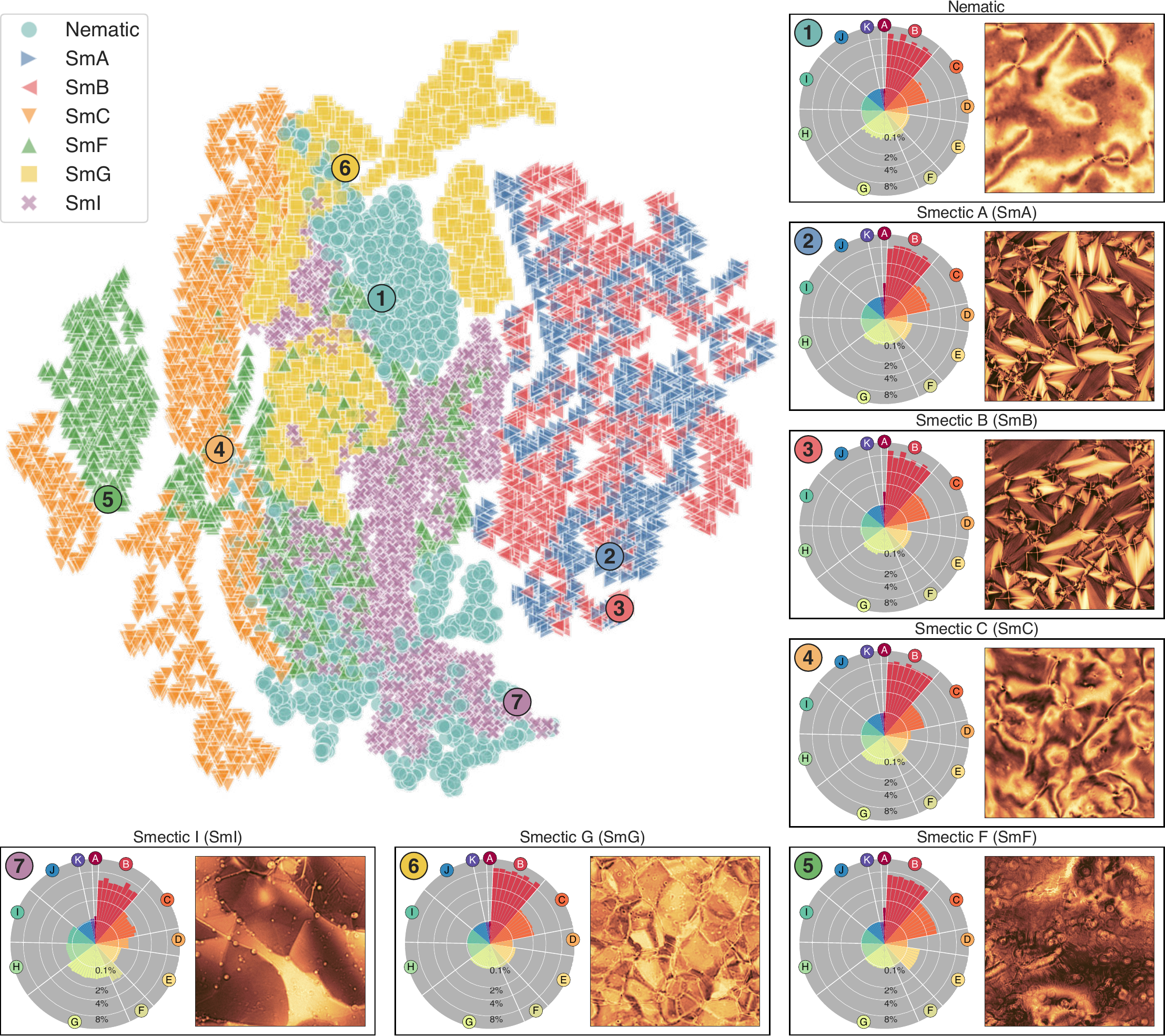}
    \caption{Mapping the ordinal-pattern space of liquid crystal textures. The main panel depicts a UMAP projection of the ordinal-pattern space, where each dimension in the original space is the normalized probability of one of the 75 possible two-by-two ordinal patterns. Each data point represents a texture in our dataset, with mesophases encoded by distinct markers and colors. The surrounding inset panels show the ordinal distributions (circular bars on a logarithmic scale) and the corresponding texture for one randomly selected example per mesophase; their locations in the UMAP plane are indicated by numbers from 1 to 7. An interactive version is provided in Ref.~\cite{voltarelli2026umaplc}.}
    \label{fig:2}
\end{figure*}

\section*{Results}

\subsection*{Ordinal-pattern feature space}

To start exploring the structure underlying the distributions of ordinal patterns and provide an illustrative overview of the diversity of textures in our dataset, we apply the uniform manifold approximation and projection (UMAP) algorithm~\cite{mcinnes2018umap, mcinnes2018umapsoftware} to project the $75$-dimensional ordinal-pattern space into two dimensions. In short, UMAP relies on a weighted graph representation (a fuzzy simplicial complex) obtained from a dissimilarity matrix (here Euclidean distance between points) to project the data into a low-dimensional space using an approach that corresponds to a force-directed layout algorithm. UMAP projections preserve meaningful data structures~\cite{lause2024art} but do not preserve high-dimensional distances~\cite{chari2023specious}. Due to its inherent stochasticity, repeated runs of the algorithm on the same data yield different, though qualitatively similar, projections, rendering them primarily suitable for comparative visual analysis. As ordinal-pattern probabilities can vary substantially in scale, we normalize their values across textures to have zero mean and unit variance before performing the UMAP projection. Figure~\ref{fig:2} shows a UMAP projection of the ordinal-pattern feature space, with colors and markers denoting the seven mesophases. The figure also displays the ordinal-pattern distributions and texture thumbnails for one randomly selected sample from each mesophase (whose locations in the projected space are indicated by numbers from 1 to 7) to illustrate their appearance under the luminance transformation. An interactive version of this figure is provided in Ref.~\cite{voltarelli2026umaplc}, enabling detailed exploration of the projected ordinal-pattern feature space.

\begin{figure*}[!ht]
    \centering
    \includegraphics[width=1\linewidth]{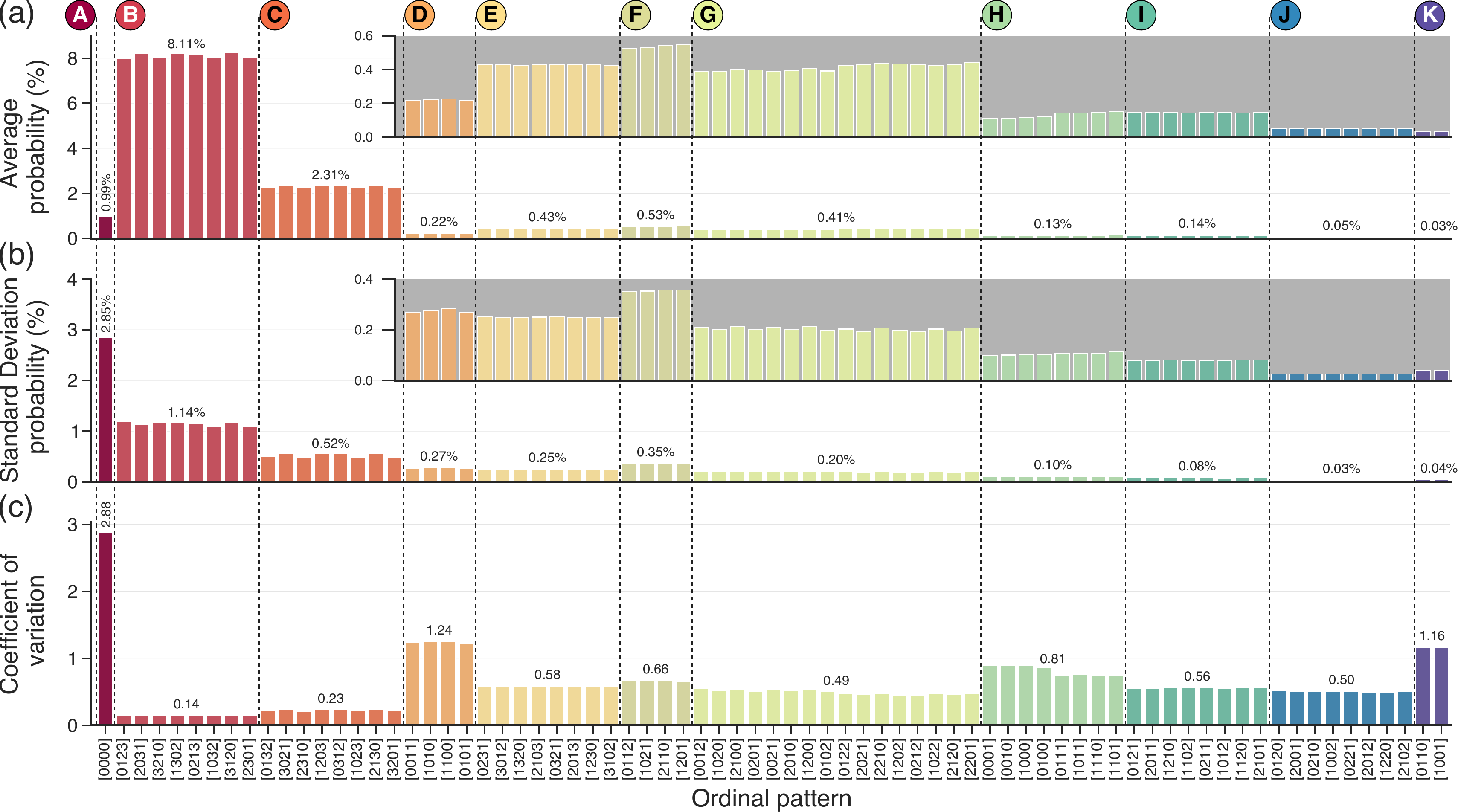}
    \caption{Overall prevalence of ordinal patterns across liquid crystal textures. Bars show the (a) average, (b) the standard deviation, and (c) the coefficient of variation (standard deviation divided by the mean) of the probability of each two-by-two ordinal pattern, calculated across all images. Patterns are arranged according to their 11 groups, indicated by colors and separated by vertical dashed lines labeled at the top. Numbers above bars denote the mean within each pattern group, and the insets in panels (a) and (b) magnify probabilities for groups D to K.}
    \label{fig:3}
\end{figure*}

Despite the substantial reduction of dimensionality in the UMAP projection, textures from the same mesophase tend to cluster together, indicating that ordinal-pattern distributions capture phase-specific textural characteristics. Nematic textures form two main clusters along the vertical midline: an upper cluster composed predominantly of classical Schlieren textures and a lower cluster comprising mainly marble and thread-like -- visual patterns that, while belonging to the same mesophase, are distinct. Textures from the SmA and SmB mesophases overlap and are concentrated on the right side of the projection; those in the upper region resemble fan-shaped patterns, while those in the lower region resemble typical focal-conic patterns, both marked by sharp lines and abrupt intensity gradients. Although transitions between these mesophases are discontinuous~\cite{dierking2003textures, dierking2025liquid}, SmA and SmB exhibit very similar textures~\cite{osiecka2023distinguishing}. This renders their ordinal-pattern distributions similar enough to appear close in this two-dimensional projection, but does not imply that they are indistinguishable in terms of these distributions; indeed, a three-component UMAP can separate SmA and SmB into distinct regions (Figure~S2 in the Supplemental Material~\cite{SI}). SmC textures form three clusters on the left side of the projection, predominantly exhibiting mosaic-like patterns with diffuse boundaries, with groups that appear to reflect different levels of uniformity (defect density within the image). The typical mosaic textures of the SmF phase are organized into two groups, one in the center of the projection and another on the left side, which, similar to the SmC case, reflect distinct levels of uniformity. SmG textures mainly exhibit well-defined mosaic-like structures and are separated into four clusters, one near the center and three in the upper-central region, consistent with variations in domain size. Finally, SmI textures appear mainly in the central–lower region and comprise both well-defined mosaic-like patterns (closer to the center) and more diffuse mosaic-like patterns (closer to the projection bottom) patterns. We observe a comparable organization when projecting the $11$-dimensional representation of textures given by the probabilities of finding patterns from groups A to K (Figure~S3 in the Supplemental Material~\cite{SI}).

\subsection*{Overall and phase-specific prevalence of ordinal patterns}

\begin{figure*}
    \centering
    \includegraphics[width=1\linewidth]{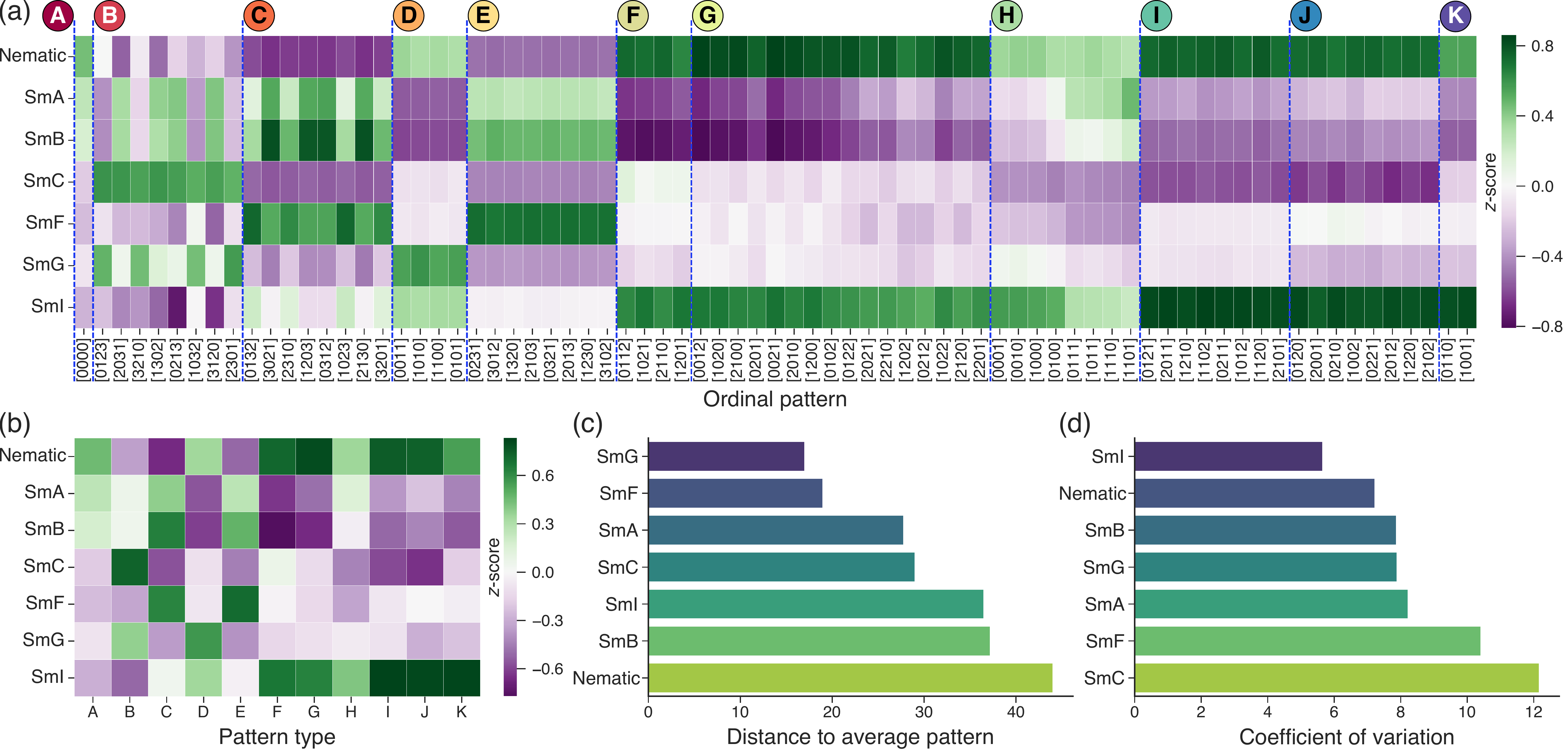}
    \caption{Ordinal-pattern fingerprints of each liquid crystal mesophase. (a) Matrix plot depicts the $z$-score probability of each two-by-two ordinal pattern (columns) across all mesophases (rows). Patterns are grouped by type and separated by vertical dashed lines. (b) Matrix plot of the $z$-score probability calculated for each pattern type (columns) and mesophase (rows). For both panels, $z$-scores are calculated by subtracting the average probability of a pattern or pattern type within a particular phase from the overall average probability and dividing the result by the overall standard deviation across all images. Positive values (green shades) indicate patterns or types that occur more frequently than the overall average, whereas negative values (purple shades) denote those that occur less frequently. Bar plots show (c) the distance to the average pattern, determined by the absolute sum of the $z$-score probabilities for each phase, and (d) the mean absolute coefficient of variation of the ordinal probabilities (in $z$-score units) calculated across all images within each mesophase.}
    \label{fig:4}
\end{figure*}

We quantify global trends in ordinal-pattern occurrence by calculating, for each pattern, the average, standard deviation, and coefficient of variation of its relative frequency across all textures. Figure~\ref{fig:3}(a) shows that average prevalence is similar within groups but differs significantly across them. Patterns from group B are the most prevalent, followed by groups C and A, whereas the remaining types (F, E, G, D, I, H, J, and K, in order of prevalence) occur at substantially lower frequencies. This behavior contrasts with that observed in art paintings, where average probabilities decrease monotonically across alphabetically arranged types~\cite{tarozo2025two} -- likely reflecting the greater diversity of patterns in paintings compared with liquid crystal textures. In turn, Figure~\ref{fig:3}(b) indicates that variability is not directly correlated with prevalence since the standard deviation decreases monotonically across types (except for type F). Figure~\ref{fig:3}(c) shows that the ranking of patterns by the coefficient of variation (standard deviation divided by the mean) differs remarkably from their ranking by average prevalence. For example, the most prevalent patterns (groups B and C) have the lowest coefficients of variation, whereas the least prevalent patterns (type K) exhibit the third-highest. These results indicate that even low-prevalence patterns can carry discriminatory information for mesophase classification and motivate normalizing raw prevalence values, since small absolute differences among low-prevalence patterns translate into significant standardized variations for several types. 

Therefore, to identify the ordinal fingerprint of each mesophase, we examine how the prevalence of every two-by-two pattern within a phase deviates from its overall prevalence by calculating the $z$-score probability for each ordinal pattern, $z_{\phi,i}= (\mathbb{E}_{\phi}[p_i] - \mathbb{E}[p_i])/\mathbb{S}[p_i]$, where $p_i$ is the probability of the $i$-th pattern, $\mathbb{E}_{\phi}[p_i]$ the mean probability of $p_i$ within the phase $\phi$, and $\mathbb{E}[p_i]$ and $\mathbb{S}[p_i]$ denote, respectively, the average probability and the standard deviation of $p_i$ throughout the dataset. Figure~\ref{fig:4}(a) shows this standardized measure of prevalence for each ordinal pattern across the seven mesophases investigated in our study. Additionally, we also calculate this $z$-score probability for the probability of finding each pattern type, as shown in Figure~\ref{fig:4}(b). These results reveal that each mesophase displays a distinct profile of pattern prevalence relative to the global average, with a group-wise organization where patterns within the same type tend to be consistently above or below the average. We emphasize that these ordinal fingerprints encode local ordering relations among neighboring pixels, capturing thus lower-level image features rather than serving as a direct proxy for eye-level texture similarity. Moreover, as shown in Figure~\ref{fig:3}, the prevalences of the most frequent types (notably B and C) vary relatively little across images, while some of the least frequent patterns (such as those in group K) exhibit much larger relative variability. Fingerprint differences between textures and phases are therefore not driven solely by the dominant patterns; instead, rare patterns can contribute disproportionately to discrimination when their relative variability is taken into account. Consequently, these ordinal fingerprints can reveal subtle distinctions that are difficult to perceive visually, while sometimes de-emphasizing global differences dominated by larger-scale morphology.

With these points in mind, we note that nematic textures show a relatively high excess of ordinal patterns with equal-intensity pixels (types A, D, F, G, H, I, J, and K), whereas patterns without equalities (types B, C, and E) have negative or near-zero $z$-scores -- consistent with the generally smooth, thread-like structures radiating from singularities typical of these textures. The focal-conic and fan-shaped textures of SmA and SmB mesophases yield similar yet distinguishable ordinal profiles: both exhibit relatively high prevalence of patterns from types A, C, and E (with types C and E further heightened in SmB) and comparatively low prevalence of patterns with two equal-intensity pixels (types D, F, G, I, J, and K). Patterns within groups B and H are not uniformly distributed as some members show positive and others negative $z$ scores -- a modulation particularly interesting for type H patterns (corner-like structures): patterns with the largest-intensity pixel at a corner ($[0111]$, $[1011]$, $[1110]$, and $[1101]$) occur more frequently in SmA than in SmB, whereas SmB shows a greater deficit of patterns with the lowest intensity pixel at a corner ($[0001]$, $[0010]$, $[1000]$, and $[0100]$) than SmA. In our dataset, SmC textures predominantly exhibit mosaic-like patterns with diffuse boundaries, and their ordinal profile is marked by negative $z$-scores across mostly patterns with identical intensities and a relatively high prevalence of type-B patterns 
-- monotonic spatial gradients along both horizontal and vertical directions -- indicating that their diffuse appearance somehow translates into more continuous gradients at the local scale. Despite also exhibiting mosaic-like textures, the ordinal profile of SmF textures roughly resembles that of SmA and SmB, with similar modulation of patterns with equal-intensity pixels; still, their distinct global visual features are reflected in marked differences in the prevalence of more common patterns from types A and B. SmG mesophase textures are characterized by well-defined mosaic-like structures that translate into a relatively high prevalence of type-D patterns (pairs of equal-intensity pixels vertically or horizontally aligned). Finally, SmI textures mirror the nematic phase in the behavior of equal-intensity patterns, while diverging for more common types such as type A (less prevalent in SmI) and types C and E (closer to their average prevalence in SmI).

We further quantify the extent to which each phase deviates from the overall distribution of ordinal patterns by summing the absolute $z$-score probabilities across all patterns, a measure that reflects how far the ordinal profile of a phase is from the global average. Figure~\ref{fig:4}(c) ranks the mesophases by this quantity. The resulting hierarchy reflects the distinct ordinal fingerprints of each mesophase, with SmG textures most closely resembling the global average, whereas the nematic phase is the most divergent. Interestingly, phases with qualitatively similar modulations of pattern prevalence (such as SmA, SmB, and SmF) are nonetheless well separated by this measure. In addition, to compare the intra-phase variability of ordinal patterns, we calculate the coefficients of variation of the $z$-score probabilities for each pattern and then average their absolute values within each phase. These averages, shown in Figure~\ref{fig:4}(d), indicate no clear association between diverging from the overall prevalence and presenting a higher dispersion in the prevalence of ordinal patterns. For instance, the nematic phase, which deviates most from the global average, has the second-lowest internal variability, whereas SmC, which occupies a middle rank in the deviation hierarchy, shows the largest variability. This within-phase dispersion provides a quantitative indication of how strongly phase heterogeneity manifests in the local ordinal statistics, reinforcing that the ordinal fingerprints summarize low-level image structure in a dataset-dependent way and do not represent universal phase invariants.

\subsection*{Machine learning mesophases}

To systematically quantify the predictive performance of the two-by-two ordinal patterns, we frame a classification task that predicts the mesophase of each texture from the prevalence of all $75$ patterns and from their aggregated prevalence across the $11$ pattern types. Although convolutional neural networks are a standard choice for image classification and have shown strong performance for liquid crystal textures~\cite{sigaki2020learning, dierking2023classification, dierking2023testing, dierking2023deep, betts2023machine, osiecka2023distinguishing, osiecka2024liquid, osiecka2024machine, betts2024possibilities, terroa2025convolutional}, we deliberately do not use deep learning approaches here because our goal is not to maximize accuracy by any means necessary, but to validate the intrinsic quality of the proposed ordinal representation. Instead, we consider the $k$-nearest neighbors algorithm~\cite{cover1967nearest}, one of the simplest machine learning classifiers~\cite{hastie2009elements}, ensuring thus that the performance primarily reflects the ordinal-pattern representation rather than model expressiveness. This instance-based method assigns unlabeled observations to the most common phase among the $k$ nearest training samples in the feature space under a specified distance metric. Because our predictive features are probabilities, we adopt the Hellinger distance~\cite{lecam1990asymptotics} (equivalent to Euclidean distance after a square-root transform of the pattern prevalences) to define local neighborhoods in the ordinal-pattern space (using Euclidean distance yields only a slightly lower performance, as shown in Figure~S5 in the Supplemental Material~\cite{SI}).

\begin{figure*}
    \centering
    \includegraphics[width=1\linewidth]{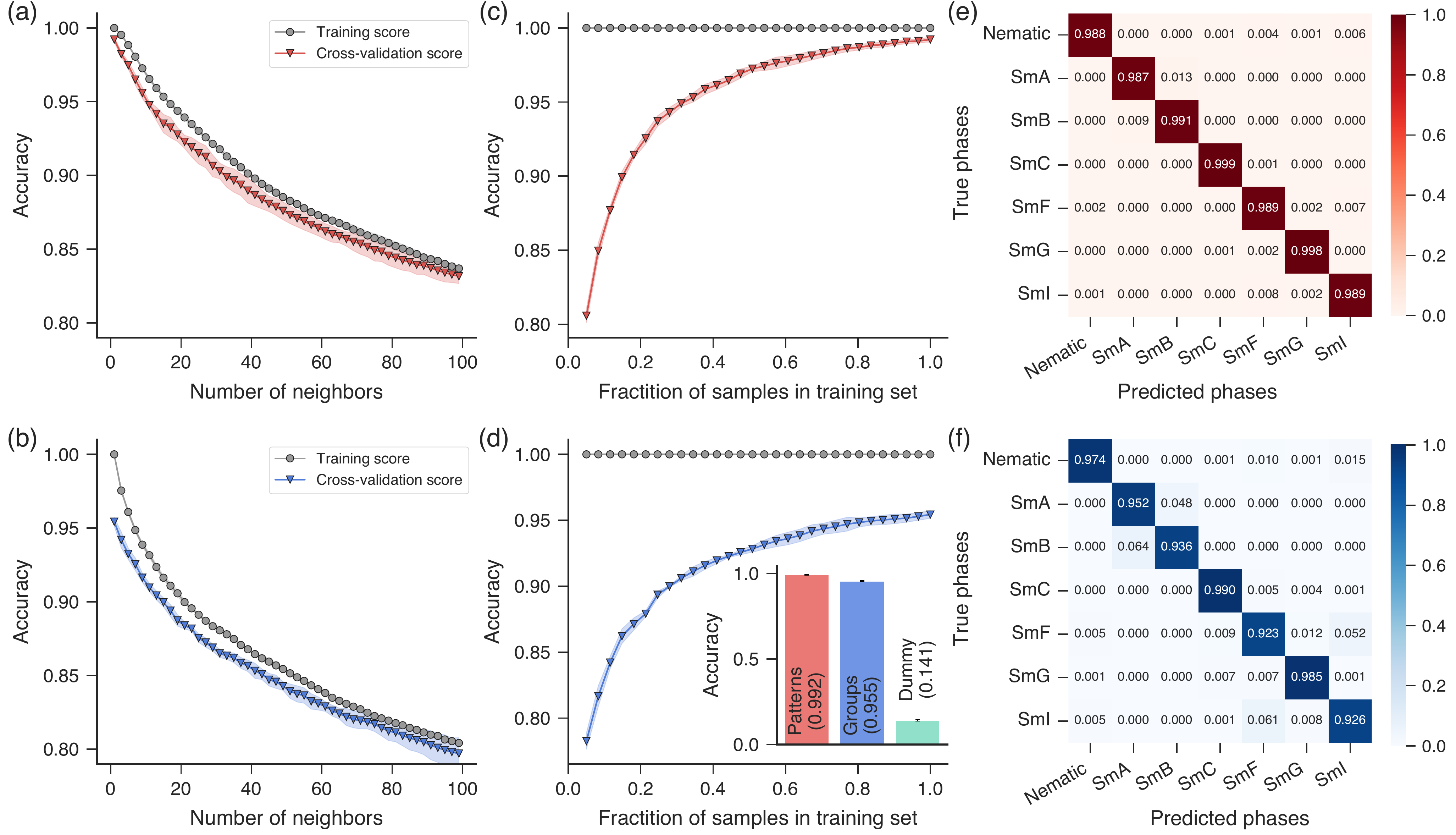}
    \caption{Classifying mesophases with two-by-two ordinal patterns. (a, b) Training (circles) and cross-validation (triangles) accuracy as a function of the number of neighbors $k$ in the nearest-neighbor classifier. (c, d) Training (circles) and cross-validation (triangles) accuracy as a function of the fraction of data used for training the nearest-neighbor algorithm with $k=1$. Shaded regions in panels (a-d) represent one standard deviation band from a five-fold cross-validation procedure. (e, f) Confusion matrices showing the fraction of classifications on the test set for each true (rows) and predicted (columns) phase. Upper panels (a, c, and e) use the prevalences of the 75 two-by-two ordinal patterns as features; lower panels (b, d, and f) use the prevalences of the 11 pattern types. The inset in (d) compares the average accuracy obtained with the 75 patterns (red) and the 11 pattern types (blue) to a dummy classifier that predicts mesophases uniformly at random (green). Results in (e), (f), and the inset are averages over 50 independent train–test splits with 80\% of the data used for training and $k=1$; error bars in the inset denote one standard deviation.}
    \label{fig:5}
\end{figure*}

We first evaluate classifier performance as a function of the number of nearest neighbors ($k$) by computing validation curves using standard five-fold cross-validation over the full dataset, whereby the data is randomly partitioned into five approximately equal subsets. In each fold, one subset serves as the validation set while the remaining four subsets are used for training. We calculate the accuracy (fraction of correct phase classifications) for each fold in the training and validation sets, averaging these values across folds and calculating their standard deviations to obtain reliable estimates for each value of $k$ between $1$ and $100$. Figure~\ref{fig:5}(a) depicts the accuracy dependence on $k$ when using all $75$ ordinal patterns as predictors, whereas Figure~\ref{fig:5}(b) shows the same dependence when training the algorithm with the prevalence of the $11$ ordinal-pattern types. In both cases, the highest accuracy scores occur at $k=1$ -- that is, textures in the validation set are classified according to the phase of their single nearest neighbor in the feature space -- and then decrease monotonically with increasing $k$. Accuracy variability in the validation sets across the five folds is minimal, as evidenced by the narrow standard deviation bands in these plots. To further assess the robustness of this finding to data partitioning, we also compute validation curves for $50$ independent random subsamples containing 80\% of the dataset, performing five-fold cross-validation within each subsample (Figure~S6 in the Supplemental Material~\cite{SI}). Across all realizations, $k=1$ consistently yields the highest cross-validation accuracy, confirming that the optimal choice $k=1$ is robust to data subsampling. The high cross-validation accuracy scores of $0.99$ using all patterns and $0.96$ using the pattern types, combined with the optimal performance at $k=1$, indicate that liquid crystal textures form well-separated mesophase clusters in the ordinal pattern space, demonstrating that these features capture phase-specific visual signatures. The higher performance achieved using all $75$ ordinal patterns shows that, despite the overall homogeneity of prevalence across pattern types, individual patterns carry additional information about mesophases.

We next examine how the performance of the $k$-nearest neighbors classifier with $k=1$ scales with training set size using learning curves. We evaluate training and validation accuracy as a function of the fraction of data allocated to training, assessing how performance evolves as more data becomes available and whether the model has reached its capacity. To generate these curves, we use the same five-fold cross-validation strategy as before, systematically varying the fraction of each training fold from 5\% to 100\% and computing the mean and standard deviation of accuracy for both training and validation sets. Figure~\ref{fig:5}(c) shows the training and validation accuracy scores as a function of the training fraction when using all $75$ ordinal patterns as predictors, whereas Figure~\ref{fig:5}(d) presents the corresponding analysis using the prevalence of the $11$ ordinal-pattern types. In both cases, training scores remain at $1$ across all training fractions (as expected for a 1-nearest neighbor classifier, since each training sample is its own nearest neighbor), while validation scores increase monotonically with diminishing returns approaching final values near $0.99$ and $0.96$ for all patterns and types, respectively. 

\begin{figure*}
    \centering
    \includegraphics[width=1\linewidth]{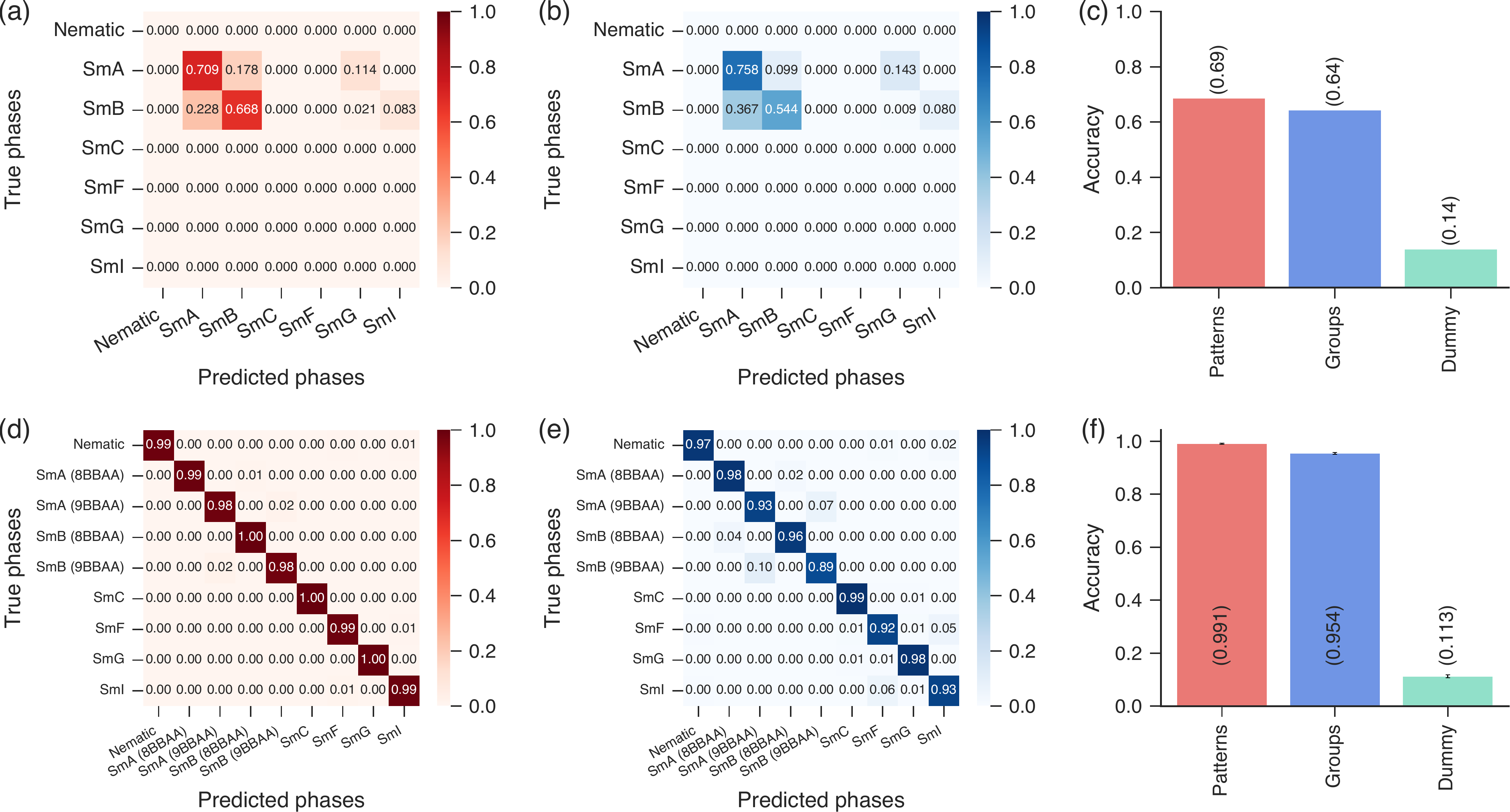}
    \caption{Classifying SmA and SmB mesophases from different compounds. (a, b) Confusion matrices showing the fraction of classifications on the set of textures of the 8BBAA Schiff-base compound, using (a) the 75 two-by-two ordinal patterns and (b) the 11 pattern types as features. In both cases, a nearest-neighbor algorithm (with $k=1$) is trained on all data except the SmA and SmB textures of 8BBAA. (c) Classification accuracy of the SmA and SmB textures of 8BBAA obtained with the 75 patterns (red) and the 11 pattern types (blue), compared with a dummy classifier that predicts mesophases uniformly at random (green). (d, e) Confusion matrices showing the fraction of classifications on the test set for each true (rows) and predicted (columns) phase (and by compound for SmA and SmB), using (d) the 75 two-by-two ordinal patterns and (e) the 11 pattern types as features. (f) Average accuracy obtained with the 75 patterns (red) and the 11 pattern types (blue), compared with the uniform-random baseline (dummy, green). Results in (d), (e), and (f) are averages over 50 independent train–test splits with 80\% of the data used for training and $k=1$; error bars in panel (f) denote one standard deviation.}
    \label{fig:6}
\end{figure*}

Additionally, we evaluate classifier performance (again fixing $k=1$) by investigating confusion matrices obtained from 50 independent train-test splits, with 80\% of the data allocated to training and 20\% to testing. Figure~\ref{fig:5}(e) presents the averaged confusion matrix when using all 75 ordinal patterns as predictors, whereas Figure~\ref{fig:5}(f) shows the corresponding results for the 11 ordinal-pattern types. In both cases, the confusion matrices exhibit a clear diagonal structure, with most entries close $0.99$ for all patterns and above $0.93$ for pattern types, indicating strong phase-specific classification. The inset of Figure~\ref{fig:5}(d) displays the average overall accuracy as a bar plot, compared against a dummy classifier that makes random predictions; the results show average accuracies of $0.99$ using all patterns and $0.96$ using the 11 types, substantially above the random baseline, and with negligible variability across the independent train-test splits (as indicated by the tiny standard-deviation error bars). When using all patterns, residual confusions remain at the $\approx$1\% level and are essentially restricted to the pairs nematic/SmI, SmA/SmB, and SmF/SmI. For the pattern types, the same pairs present slightly larger confusion fractions: SmA/SmB and SmF/SmI are confused $\approx$5\% of the time, while nematic and SmI are confused with each other $\approx$2\% of the time. The reduced performance observed when using the pattern types mainly comes from these three pairs, indicating that individual ordinal patterns carry additional information that is particularly relevant for discriminating among these mesophases. Although it is difficult to associate these residual misclassifications with global visual features of the textures, we note that nematic and SmI textures overlap primarily in the bottom part of the UMAP projection in Figure~\ref{fig:2}, a region that mainly contains nematic textures with marble and thread-like patterns and SmI textures with more diffuse mosaic-like patterns. Similarly, the confusion between SmF and SmI likely reflects their shared mosaic-like appearance, whereas the confusion between SmA and SmB is consistent with their well-known global visual similarity. We remark that similarly high accuracy has also been reported with deep learning methods in related contexts involving mesophase identification from liquid crystal textures~\cite{sigaki2020learning, dierking2023classification, dierking2023testing, dierking2023deep, betts2023machine, osiecka2023distinguishing, osiecka2024liquid, osiecka2024machine, betts2024possibilities, terroa2025convolutional}. Our approach, however, is lightweight and relies on a transparent representation. As we discuss in the next section, these properties enable a favorable trade-off among accuracy, simplicity, and interpretability, which is difficult to achieve with deep learning models.

To test generalization to unseen compounds, we focus on distinguishing SmA from SmB using textures from the 8BBAA and 9BBAA compounds due to their highly visual similarity~\cite{osiecka2023distinguishing}. Because 8BBAA and 9BBAA are adjacent homologues, they are expected to exhibit very similar textures, behavior, and properties; nevertheless, training on one compound and testing on the other probes how well phase-specific ordinal signatures transfer across materials. Thus, we train the $k$-nearest neighbors classifier fixing $k=1$ and excluding all SmA and SmB textures from the 8BBAA Schiff-base compound (retaining only those from 9BBAA in the training set), and then test performance on textures from 8BBAA, thereby assessing whether the learned patterns transfer across materials. Figure~\ref{fig:6}(a) presents the confusion matrix using all $75$ ordinal patterns, whereas Figure~\ref{fig:6}(b) shows the results for the $11$ pattern types. The overall accuracy, displayed as a bar plot in Figure~\ref{fig:6}(c), is $0.69$ for all patterns and slightly lower at $0.64$ for pattern types, values substantially higher than the random baseline of $0.14$ but significantly lower than those obtained when SmA and SmB textures from both compounds are pooled for training and testing. In both cases, the algorithm mainly confuses SmA and SmB with each other, with SmB more frequently predicted as SmA (23\% of the time with all patterns and 37\% with pattern types). Another source of confusion arises when SmA is misclassified as SmG and when SmB is mistaken for the SmI phase, likely reflecting the similarities of some mosaic patterns with focal conic domains. Although less accurate than the previous classification task, these results indicate a degree of generalization, showing that the prevalence of patterns and types learned from the 9BBAA compound transfers to the 8BBAA material only partially. At the same time, the decrease in accuracy despite strong visual similarity suggests that the ordinal-pattern representation may encode compound-specific structure within the same phase. Figure~S4 in the Supplemental Material~\cite{SI} shows analogous results when training the model excluding textures from 9BBAA, where a similar behavior is observed, albeit with slightly lower overall accuracy and with most confusion occurring between the SmA and SmB phases. These results are consistent with the fact that ordinal pattern capture very local image structures, so textures that appear highly similar at the global scale can still yield distinct ordinal pattern distributions.

Finally, we test for compound-specific signatures in the ordinal-pattern representation. We split SmA and SmB textures by compound, assigning distinct labels to SmA from 8BBAA, SmA from 9BBAA, SmB from 8BBAA, and SmB from 9BBAA, along with labels for all other mesophases. Once again, we allocate 80\% of the data for training the $k$-nearest neighbors classifier with $k=1$ and test on the remaining 20\%, further replicating the train-test split 50 times and calculating average confusion matrices using all 75 ordinal patterns and the 11 pattern types as predictors. Figure~\ref{fig:6}(d) presents the results for all patterns, whereas Figure~\ref{fig:6}(e) shows those for the 11 pattern types. The average overall accuracy, displayed in Figure~\ref{fig:6}(f), is $0.99$ for all patterns and $0.95$ for pattern types, both significantly higher than the random baseline and similar to those observed when aggregating textures from both compounds without distinguishing their origin. Most residual confusion occurs between SmB and SmA from the 9BBAA compound ($\approx$2\% with all patterns and $\approx$10\% with pattern types), while other off-diagonal entries are negligible. This finding is consistent with the UMAP projection of the ordinal pattern representation shown in Figure S2 in the Supplemental Material~\cite{SI}, where SmA and SmB textures also form well-separated clusters by compound. Thus, when the classifier learns these compound-specific features alongside phase information, discrimination becomes easier as textures separate along two dimensions: mesophase identity and material origin. This near-perfect compound-aware performance supports the interpretation that, beyond phase information, our ordinal pattern also retains subtle compound-specific signatures within SmA and SmB mesophases. In contrast, training on one compound while testing on another (as in the previous task) requires the model to rely solely on phase-specific features that generalize across materials, a considerably more challenging problem because compound-specific variations act as confounding factors that obscure the phase-distinguishing signatures. Therefore, the same locality that helps reveal subtle compound-specific structure can also reduce cross-compound transferability.

\subsection*{Interpreting ordinal pattern contributions}

To elucidate how ordinal-pattern types contribute to phase classification, we apply the SHAP (SHapley Additive exPlanations) framework~\cite{lundberg2017unified}, a model-agnostic method based on cooperative game theory that quantifies each feature's contribution to individual predictions. We use this framework for its ability to provide consistent and locally accurate contributions for both individual and interaction effects. Because SHAP calculations for $k$-nearest neighbors classifier require model-agnostic explainers (such as the KernelSHAP~\cite{lundberg2017unified}) that rely on repeated sampling of feature subsets and are unmanageable at our dataset size, we replace it with the XGBoost algorithm~\cite{chen2016xgboost}, a gradient-boosted decision tree model that enables efficient SHAP computation via the TreeSHAP algorithm~\cite{lundberg2020local2global}. We use XGBoost here solely as an explainable surrogate model trained on the same ordinal features, thereby enabling us to quantify the contributions of individual ordinal pattern types and their interactions in a practical way. Moreover, despite the greater model complexity of XGBoost compared to the nearest neighbors method, Figure~S7 in the Supplemental Material~\cite{SI} shows that both classifiers yield very similar accuracies across all mesophases, corroborating the quality of our ordinal pattern representation. For interpretability, we focus on the prevalence of the $11$ ordinal pattern types rather than all $75$ individual patterns, retaining essential texture characteristics that enable accurate phase classification.

\begin{figure*}
    \centering
    \includegraphics[width=1\linewidth]{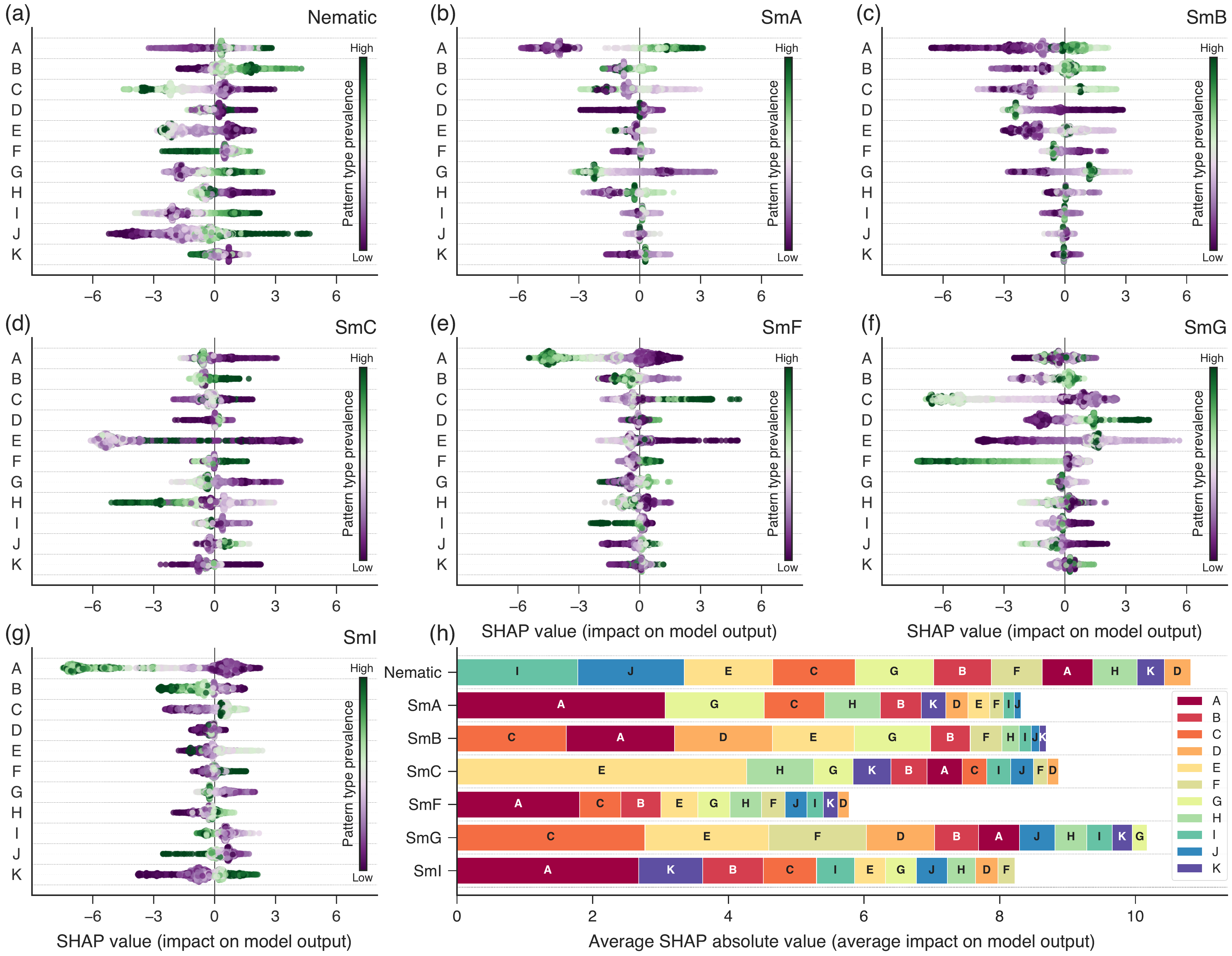}
    \caption{Interpreting the contribution of ordinal pattern types to mesophase classification. Panels (a)-(g) show density scatter plots of SHAP values for each ordinal pattern type across mesophases. SHAP values ($x$-axis) quantify the impact of ordinal pattern type on the model output for individual observations. Points are color-coded by the prevalence of the pattern type in each observation, with green denoting high prevalence and purple denoting low. Panel (h) displays horizontally stacked bars of the mean absolute SHAP value for each pattern type across mesophases. Within each mesophase, pattern types are ordered by decreasing mean SHAP magnitude, with higher values indicating types with greater influence on the model's predictions.}
    \label{fig:7}
\end{figure*}

We train the XGBoost classifier using pattern type prevalences as predictors and compute SHAP values for each prediction in the test set. These values quantify the contribution of each pattern type to the log-odds of membership in a given mesophase, with positive values increasing and negative values decreasing the probability of phase membership. Figures~\ref{fig:7}(a)–(g) present density scatter plots of SHAP values for each ordinal pattern type across the seven mesophases, with colors indicating the corresponding pattern type prevalence. Figure~\ref{fig:7}(h) reports the mean absolute SHAP value for each pattern type across all mesophases, revealing that types A, B, C, and E often rank among the most informative features for phase classification. 

The model uses multiple distinctive patterns to discriminate textures, and we highlight a few key associations. For the nematic phase, the classifier primarily relies on types I, J, and E, with increasing prevalence of I and J raising the log-odds of nematic membership, while higher prevalence of E tends to diminish it. For SmA and SmB, type A prevalence positively correlates with membership in both phases. The increasing prevalence of type C is associated with SmB membership, whereas the association is less clear for SmA. Notably, type G prevalence shows opposing effects -- negatively correlated with SmA and positively correlated with SmB membership -- providing a discriminative signature between these visually similar textures. For SmC, the model relies most on type E (negatively correlated with phase membership), followed by type H. Type A predominantly supports the identification of the SmF phase, with higher prevalence decreasing the log-odds of belonging to this phase. The model primarily uses types C, E, and F to identify the SmG phase, with the prevalence of C and F negatively correlated and E positively correlated with phase membership. For SmI, the classifier mainly relies on type A, whose prevalence negatively associates with the log-odds of belonging to this phase; types K, B, and C are also frequently used, with K and C positively correlated and B negatively correlated with phase membership.

We move beyond individual contributions to investigate the role of interactions among pattern types in the model output. SHAP interaction values extend the standard SHAP framework by quantifying how the contribution of one feature depends on the value of another feature~\cite{lundberg2018consistent, lundberg2020local2global}. These values form a matrix for each prediction, where diagonal elements represent main effects and off-diagonal elements capture interaction effects after accounting for individual contributions. To visualize individual and interaction effects simultaneously, we adopt a recently proposed network-based approach~\cite{furger2025single} to create phase-specific interaction networks. 

\begin{figure*}
    \centering
    \includegraphics[width=1\linewidth]{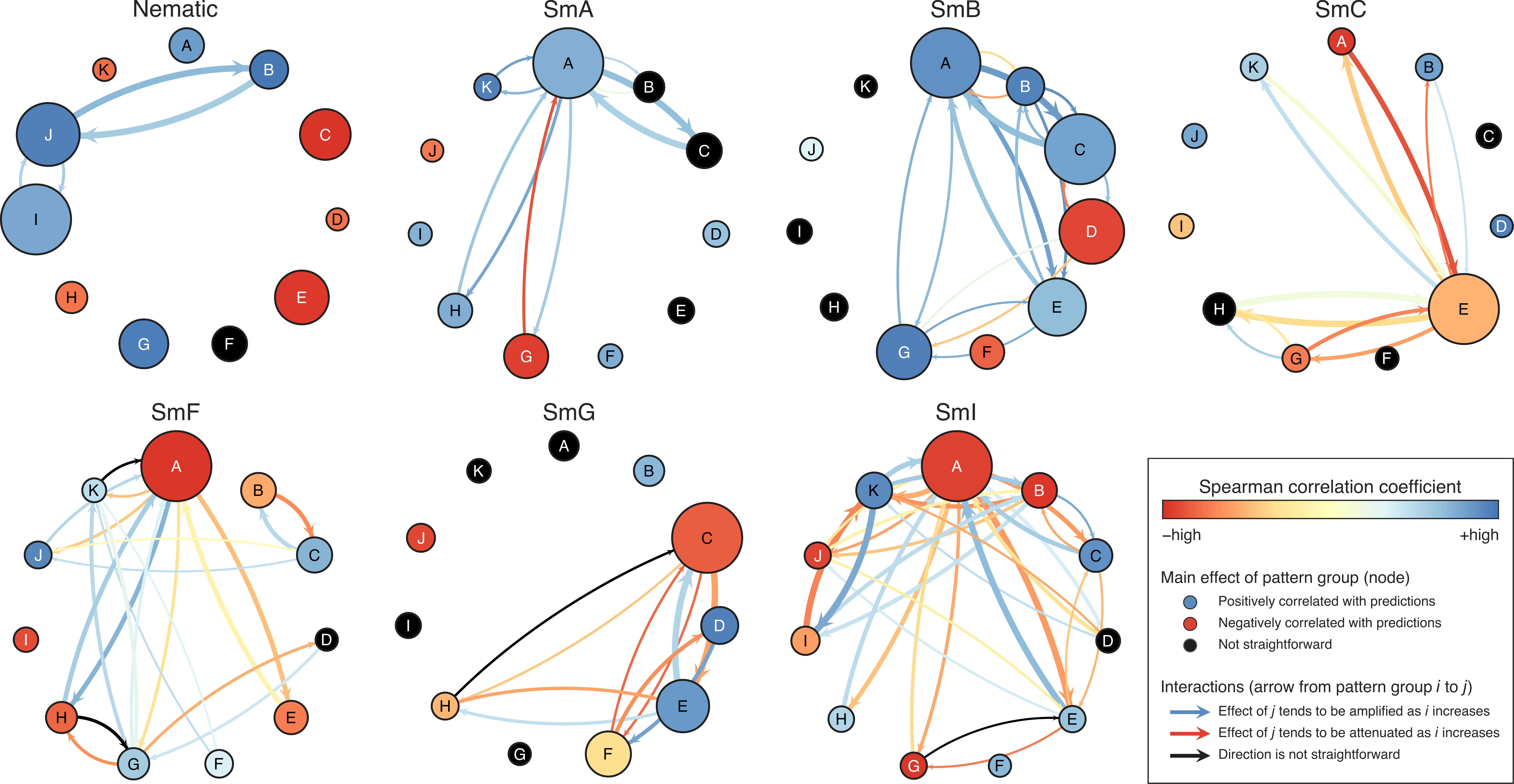}
    \caption{Interaction among ordinal-pattern types in mesophase classification. Each graph represents individual effects of ordinal-pattern types and their pairwise interactions on model predictions for a given mesophase. Nodes represent pattern types, with sizes proportional to the mean absolute SHAP value, and colors encoding the Spearman correlation between SHAP values and the prevalence of pattern type. Blue tones indicate a positive association with mesophase membership, and red tones indicate a negative association. Directed edges signify that the SHAP interaction for a pair of pattern types correlates with the prevalence of the source pattern type. Blue edges denote amplification, meaning higher prevalence of the source pattern type increases the impact of the target pattern type. Red edges denote attenuation, meaning higher prevalence of the source pattern type decreases the impact of the target pattern type. Only edges with an absolute Spearman correlation of at least $0.3$ are shown. Black nodes and edges flag ambiguous directionality, where Spearman and Pearson correlations have opposite signs.}
    \label{fig:8}
\end{figure*}

Figure~\ref{fig:8} presents the interaction networks for each mesophase, with nodes corresponding to the 11 pattern types. Node size denotes the mean absolute SHAP value (overall importance for model predictions), and node color encodes the Spearman correlation coefficient between SHAP values and pattern prevalence (blue tones indicate positive correlations; red tones indicate negative correlations). Directed edges represent interactions between pattern types and are drawn when the Spearman correlation between SHAP interaction values and the prevalence of the source node is statistically significant. Positive correlations (blue shades) indicate that a higher prevalence of the source pattern enhances the impact of the target pattern on the model output; negative correlations (red shades) indicate attenuation of that effect. Each phase exhibits a fingerprint-like network that the model uses for classification. Nodes and edges in black flag cases where Spearman and Pearson correlation coefficients have different signs, suggesting that the direction of association is not well defined. Additionally, only edges with absolute Spearman correlation values higher than $0.3$ are shown. 

Among the many interaction patterns, we highlight a few representative examples. In the nematic phase, types J and B show positive synergy: increasing the prevalence of both amplifies their positive main effects on nematic membership. In SmA, the pairs A-K and A-H exhibit positive synergy, whereas in SmB, several other pairs (such as A-C, A-E, A-G) are synergistic, but A-K and A-H are not even related. In SmC, a mutual attenuation of negative effects on the model output occurs for pairs A-E and E-G. In SmF, the interaction between A and H shows negative synergy: decreasing the prevalence of both amplifies their negative main effects on SmF membership. In SmG, the interaction between types C and E is asymmetric: higher prevalence of C attenuates the positive effect of E, whereas higher prevalence of E enhances the negative effect of C. SmI is the most complex case, with numerous non-synergistic interactions. These network representations reveal the intricate phase-specific interplay among the prevalence of pattern types that underlie accurate mesophase classification.

\section*{Discussion and conclusions}

Our study has thus demonstrated that ordinal pattern representations of liquid crystal textures from a large-scale dataset provide a robust and interpretable feature space for classifying mesophases. Integrating this representation with a very simple machine learning classifier has proven enough to achieve near-perfect classification accuracy across seven distinct mesophases, including the challenging distinction between smectic A and smectic B phases, which are characterized by highly similar optical textures~\cite{osiecka2023distinguishing}. Beyond its high classification accuracy, our ordinal pattern approach offers substantial advantages in interpretability. In contrast to the often opaque nature of features automatically extracted from deep convolutional neural networks, each ordinal pattern represents simple local pixel intensity relationships that are easily conceptualized, enabling expert visual exploration and making them relatable to the physical characteristics of liquid crystal textures. Furthermore, by combining the SHAP framework~\cite{lundberg2017unified, lundberg2018consistent, lundberg2020local2global} with a novel network visualization technique~\cite{furger2025single}, we have dissected not only the individual contributions of each pattern type but also how their interactions drive the classification decisions. The resulting interaction networks offer compact and readable summaries of the complex relationships among pattern types underlying each phase classification -- a tool potentially useful for deepening our understanding of the textural characteristics that define each mesophase.

We have also verified the capability of the ordinal pattern representation to generalize to textures from unseen compounds. Although accuracy decreased when the model was trained on one compound and tested on another for classification of smectic A and smectic B phases, performance remained significantly above a random baseline, indicating that the ordinal pattern signatures for mesophases are, to some extent, material-agnostic. Conversely, when the classifier was trained to distinguish phase identity and material origin simultaneously, the same approach yielded almost perfect discrimination. These results indicate that the prevalence of ordinal patterns encodes both phase identity and material-specific characteristics, opening possibilities for applications in material identification and characterization of compositional variations.

Nevertheless, our work is not without its limitations. Cross-compound generalization, while significant, is far from perfect, indicating that compound-specific features act as confounders that obscure phase signatures. The smectic A and smectic B phases analyzed here correspond specifically to the transition from fluid smectic A to soft-crystal B~\cite{osiecka2023distinguishing}, rather than the more subtle and challenging transition from fluid smectic A to hexatic smectic B discussed in Ref.~\cite{betts2024possibilities}. Because the compounds investigated here do not exhibit a hexatic smectic B phase in their polymorphism, our results should be interpreted in the context of the transition from fluid smectic A to soft-crystal B, and testing our ordinal pattern framework on datasets that include a hexatic smectic B phase remains an important and more stringent direction for future work. More broadly, although our results cover seven distinct mesophases and show consistent performance across the phase boundaries within this set, they include only a limited subset of soft-crystal phases and transitions (beyond the soft-crystal B, only the smectic G phase is represented). Consequently, expanding the dataset to include a broader range of soft-crystal phases and transitions -- such as the crystalline G-crystalline H, crystalline J-crystalline K, and crystalline B-crystalline E transitions -- is a key next step toward establishing the generality of our ordinal pattern framework. In addition, larger datasets spanning other emergent textures (such as twist-grain boundary, columnar, and blue phases), as well as different confining geometries, would help delineate the scope and limits of our findings. Moreover, the ordinal pattern representation relies on grayscale-transformed textures,  which confers robustness to varying imaging conditions but implies a loss of potentially helpful information. Future research could thus explore multiscale, color- and symmetry-aware ordinal patterns that may further improve the strength of our approach and aid in the search for fully universal mesophase characteristics. Another possibility is to account for temporal evolution under temperature changes to probe dynamic features of textures and the impact of the phase sequence history of each sample on classification performance. Finally, extending our ordinal pattern framework to other complex patterned systems in materials science, biology, and geology represents a promising avenue for future work.

\section*{Acknowledgments}
We acknowledge the support of the Coordena\c{c}\~ao de Aperfei\c{c}oamento de Pessoal de N\'ivel Superior, the Conselho Nacional de Desenvolvimento Cient\'ifico e Tecnol\'ogico (CNPq -- Grants 303533/2021-8 and 303864/2024-9), and Slovenian Research and Innovation Agency (Javna agencija za znanstvenoraziskovalno in inovacijsko dejavnost Republike Slovenije) (Grant P1-0403).
 
\section*{Data availability}
The data used in the current study are available from the corresponding author upon request.

\bibliography{references.bib}

\clearpage
\includepdf[pages=1-1,pagecommand={\thispagestyle{empty}}]{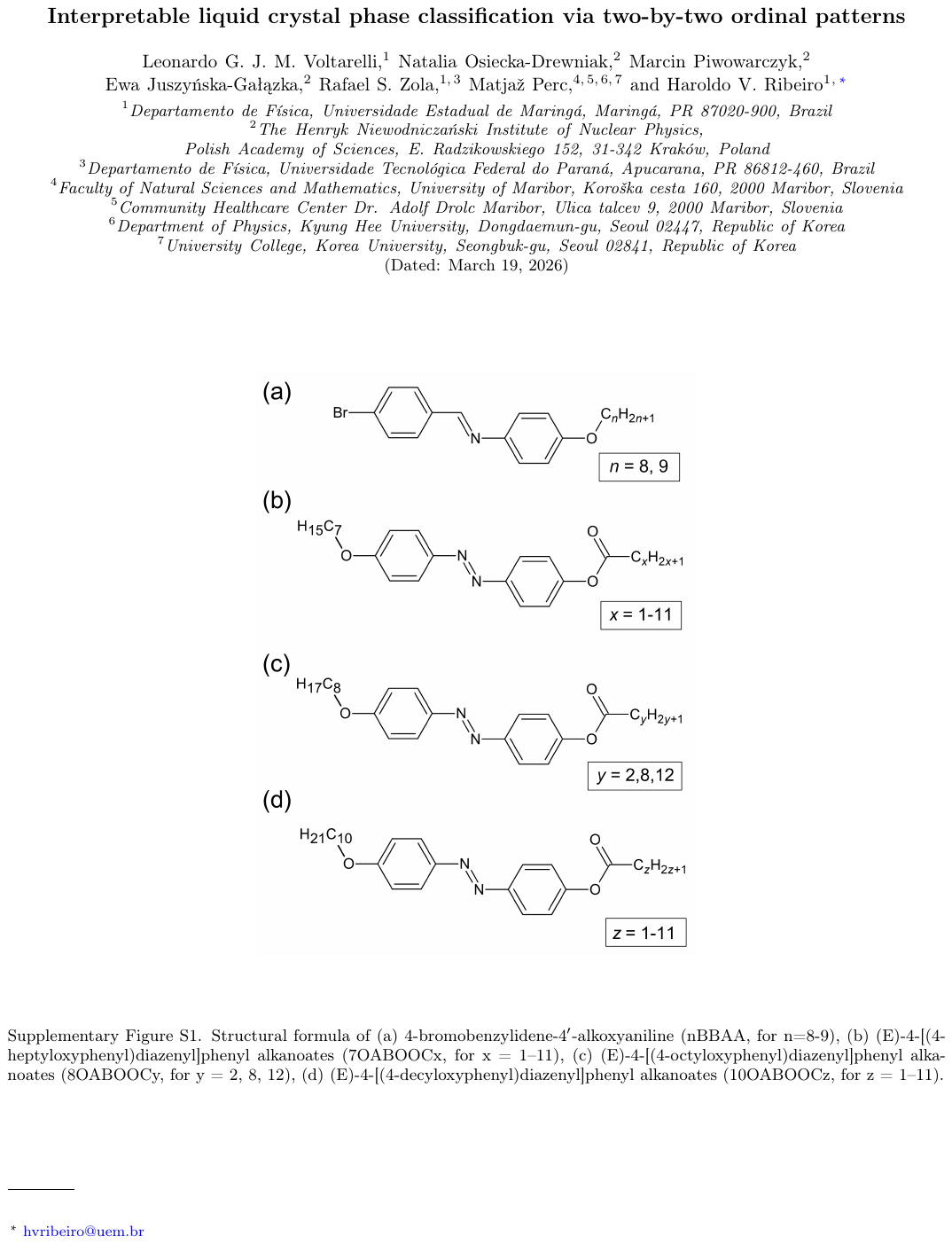}
\clearpage
\includepdf[pages=2-2,pagecommand={\thispagestyle{empty}}]{supplementary.pdf}
\clearpage
\includepdf[pages=3-3,pagecommand={\thispagestyle{empty}}]{supplementary.pdf}
\clearpage
\includepdf[pages=4-4,pagecommand={\thispagestyle{empty}}]{supplementary.pdf}
\clearpage
\includepdf[pages=5-5,pagecommand={\thispagestyle{empty}}]{supplementary.pdf}
\clearpage
\includepdf[pages=6-6,pagecommand={\thispagestyle{empty}}]{supplementary.pdf}
\clearpage
\includepdf[pages=7-7,pagecommand={\thispagestyle{empty}}]{supplementary.pdf}
\clearpage

\end{document}